\documentclass[
 twocolumn,
 amsmath,amssymb,
 aps, physrev,
 prl,
]{revtex4-2}

\usepackage{color}
\usepackage{amsmath}
\usepackage{pifont}   
\usepackage{graphicx} 
\usepackage{dcolumn}  
\usepackage{bm}       
\usepackage{amsfonts} 
\usepackage{amssymb}  
\usepackage{multirow} 
\usepackage{siunitx}
\usepackage[dvipsnames]{xcolor}
\usepackage[english]{babel}
\usepackage[autostyle, english = american]{csquotes}
\MakeOuterQuote{"}

\usepackage{placeins}

\usepackage{braket} 

    \renewcommand{\v}[1]{\bm{\mathrm{#1}}}
    \newcommand{\m}[1]{\bm{\mathsf{#1}}}

\newcommand{\mel}[3]{\left\langle #1 \left| #2 \right| #3 \right\rangle}

\begin{document}

\title{Generation of pure, spin polarized, and unpolarized charge currents at the few cycle limit of circularly polarized light}

\author{D. Gill}
\affiliation{Max-Born-Institute for Non-linear Optics and Short Pulse Spectroscopy, Max-Born Strasse 2A, 12489 Berlin, Germany}
\author{S. Sharma}
\affiliation{Max-Born-Institute for Non-linear Optics and Short Pulse Spectroscopy, Max-Born Strasse 2A, 12489 Berlin, Germany \\
Institute for theoretical solid-state physics, Freie Universit\"at Berlin, Arnimallee 14, 14195 Berlin, Germany}
\author{S. Shallcross}
\email{sam.shallcross@mbi-berlin.de}
\affiliation{Max-Born-Institute for Non-linear Optics and Short Pulse Spectroscopy, Max Born Strasse 2A, 12489 Berlin, Germany}

\date{\today}

\begin{abstract}
In certain members of the transition metal dichalcogenide (TMDC) family, laser pulses of oppositely circularly polarized light excite electrons of opposite spin. Here we show that in the few cycle limit such pulses generate not only a spin density excitation, but also a  spin current excitation. Employing the example of the TMDC WSe$_2$ we show that {\it pure} spin currents, the flow of spin in the absence of net charge flow, 100\% spin polarized currents, and charge currents are all accessible and controllable by tuning the amplitude of $\sim5$ femtosecond gap tuned light pulses. Underpinning this physics is a symmetry lowering of the valley charge excitation from $C_3$ at long duration to $C_2$ in the few cycle limit, imbuing the excitation with net current. Our results both highlight the emergence of a rich light-spin current coupling at ultrafast times in the TMDC family, as well presenting a route to the all-optical generation of pure spin currents.
\end{abstract}

\maketitle



The electron possesses, in addition to charge, a fundamental two state freedom: the electron spin. While charge currents form the basis of present day semiconductor electronics, the spin attribute offers the possibility of a flow of spin {\it without} a net charge flow\cite{hoffmann2007pure,
uchida2008observation,uchida2008thermo,
ligeneration2014,yunonlinear2014,
hugatetunable2023}. 
This fundamental physics offers new technological horizons involving e.g. reduced dissipation transport, spintronic THz emission~\cite{fulop_laser-driven_2020}, and ultrafast spin-moment control\cite{huang2020pure}. A concomitant research drive thus exists to understand, create, and control pure spin currents.

Diverse mechanisms exist by which pure spin currents (PSC) can be generated, possessing distinct realms of temporal applicability. At the longest time scales the 
spin Seebeck\cite{uchida2008observation,uchida2008thermo,
ligeneration2014,yunonlinear2014,hugatetunable2023}
and spin Hall effects\cite{pham2016ferromagnetic,
khang2018conductive} generate PSC via thermodynamic and electric potentials respectively. Excitation of PSC by light pulses offers creation on dramatically quicker time scales, which can be of the order of a few picoseconds to several hundred femtoseconds. The photo-galvanic effect generates a pure direct current from an oscillating light pulse, via the excitation of a distribution of electrons balanced between  spin channels\cite{yu_nonlinear_2014,
shan_optical_2015,xie_photogalvanic_2015,xie_two-dimensional_2018,
jin_photoinduced_2018,
mu_pure_2021,
xu_pure_2021,zhang_robust_2023}.
This requires a particular symmetry to the band structure, created either structurally -- for example edge states in nanostructures\cite{chen_photogalvanic_2018,
jiang_robust_2019,tao_pure_2020,
zhou_pure_2021,li_realizing_2023} -- or intrinsically as for example in the trigonally warped and spin split valley manifolds of WSe$_2$\cite{yu_nonlinear_2014}.

Exploiting the recent observation that few cycle circularly polarized light pulses generate valley current\cite{sharma_direct_2024,gill_creation_2024}, we here propose an ultrafast all-optical route to the generation of PSC. 
For a strong spin orbit transition metal dichalcogenide such as WSe$_2$, few cycle pulses endow the valley manifolds with a current carrying "dipole like" momentum space excitation. This dipole can, via pulse parameters, be tuned within each spin channel separately to achieve a range of current responses to light, from pure spin currents to fully polarized spin currents.
Our scheme requires no underlying material symmetry -- either intrinsic or engineered -- and allows independent control of the valley polarization (via the pulse helicity) and spin purity (via the pulse amplitude and duration). While light-control by circularly polarized  pulses over valley charge is now well established
\cite{xiao_coupled_2012,mak_control_2012,
xiao_nonlinear_2015,schaibley_valleytronics_2016}, our findings demonstrate that in the few cycle limit this  process offers control also over spin currents, providing a potential route to the  creation of pure spin current at few femtosecond time scales.

\begin{figure*}[t!]
\includegraphics[width=0.6\textwidth]{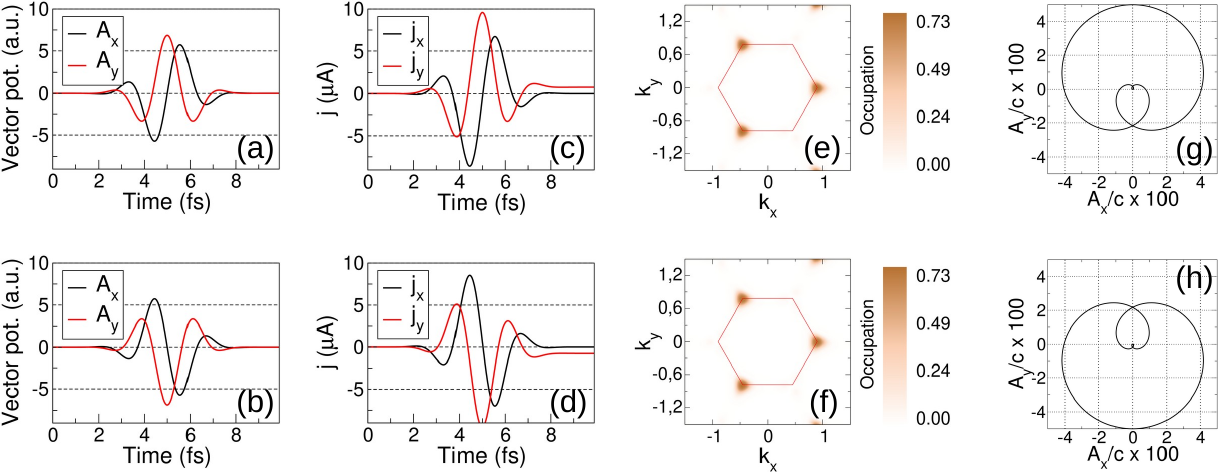}
\caption{{\it Current generation by single cycle circularly polarized light in "gapped graphene"}. The vector potential of the two single cycle pulses shown in panels (a,b) generates oppositely directed post pulse residual currents, with $j_y > 0$ and $j_y < 0$ respectively, panels (c,d). This dramatically different current response, which occurs despite very similar momentum resolved excitations, panels (e,f), represents a signature of a short time light-matter symmetry breaking. The momentum space trajectory of the two pulses shown in (a,b), panels (g,h) respectively, are seen to possess $C_2$ symmetry, lower than the $C_3$ symmetry of the valley manifold, and this is responsible for the generation of current.
}
\label{fig1}
\end{figure*}


\emph{Current response at the few cycle limit:} We first discuss the basic symmetry breaking properties of single cycle and few cycle circularly polarized light. The simplest model exhibiting this physics is that of "gapped graphene"\cite{xiao_coupled_2012,sharma_valley_2022}, a graphene lattice in which a staggered sub-lattice potential is applied to open a gap. We employ this model here, treating the dynamics of light-matter interaction via the von Neumann equation of motion with a phenomenological decoherence time of 20~fs\cite{heide_electronic_2021}. Full details of this approach can be found in the supplemental document.

Two pulses of single cycle circularly polarized light, vector potentials shown in Figs.~\ref{fig1}(a,b), generate a steady state post pulse current of opposite sign, panels (c,d), despite possessing very similar valley polarized charge excitation, panels (e,f). Note that here we show the intraband current component, the full current possesses a decaying oscillatory interband component such that it limits to the intraband component at longer times\cite{higuchi_light-field-driven_2017,heide_lightwave-controlled_2019}, see supplemental document. 
%
%
The origin of the opposite sign of the current generated by the light pulses shown in Figs.~\ref{fig1}(a,b) is revealed by plotting the dynamical trajectory, i.e. the evolution of crystal momentum $\v k(t)$ that each pulse induces, Figs.~\ref{fig1}(g,h). These are seen to represent single loops that clearly break $k_y$ mirror symmetry in momentum space, leading to the finite $j_y$ current component seen in Figs.~\ref{fig1}(c,d). A similar pulse form, created via co-circular $\omega$-$2\omega$ light pulses, has been explored in Ref.~\cite{neufeld_light-driven_2021}.

\begin{figure}[t!]
\includegraphics[width=0.48\textwidth]{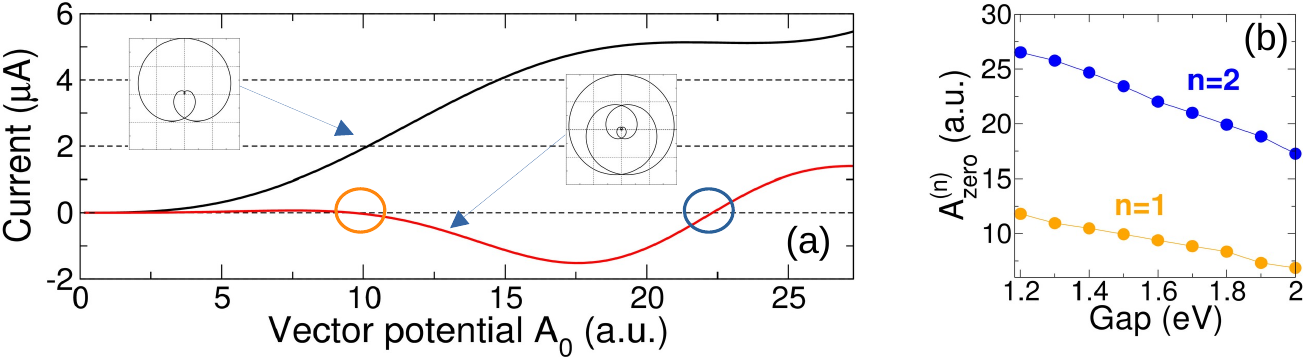}
\caption{{\it Control over current in the few cycle regime of circularly polarized light in "gapped graphene"}. (a) The dependence of current on vector potential amplitude is strikingly and qualitatively different for single cycle (black line) and a few cycle (red line) pulses: the latter exhibits zeros in the current, while the former presents only a monotonic increase with pulse amplitude. The inset panels display the momentum space trajectory (Lissajous figure) for the single and few cycle pulse respectively. The critical amplitude at which the zeros in the current occur, highlighted by the circles, depends on the gap as shown in panel (b).
}
\label{fig2}
\end{figure}

In a few cycle pulse the dynamical trajectory will consist not of a single loop, but a series of increasing loops alternating between $k_y>0$ and $k_y<0$ dominance. 
As these loops individually would generate current of opposite sign, a series of such loops in one pulse will generate cancellation effects in the post-pulse current. Such cancellation effects, that can lead to current of either sign or even zero current, will evidently depend on the pulse parameters.
To see this we plot the residual post pulse current for a few cycle pulse as a function of the vector potential amplitude, Fig.~\ref{fig2}(a), revealing a post-pulse current alternating between $j_y > 0$ and $j_y < 0$ with two values of the vector potential at which the pulse generates zero current.
For comparison we show also the case of a single cycle pulse for which, in contrast, yields only a monotonic increase in the current. Note that inset panels in Fig.~\ref{fig2}(a) represent the momentum space trajectories $\v k(t)$ (Lissajous figures) of each of these two pulses.
As the area of momentum space endowed with Berry curvature depends on the gap size, vanishing in the limit in which the gap goes to zero, the critical points at which the current changes sign will also depend on the gap. This can be seen in Fig.~\ref{fig2}(b), with an increase in the gap reducing monotonically the values of the vector poptential amplitude for which current zeros $A_{zero}^{(n)}$ occur.


\emph{Spin and charge currents at the few cycle limit:} This gap dependence of the zeros of the few cycle current response suggests that for a material such as WSe$_2$, that possesses different gaps in the spin up and spin down channels, the spin current response to few cycle light may be particularly rich. By tuning the vector potential amplitude the spin up and spin down manifolds can be endowed with current of either direction, or no current. This implies the possibility of the charge current cancelling when summed over both manifolds, i.e. of creating an ultrafast pure spin current response: a net spin flow in the absence of a net charge flow.

To explore this we consider a 4-band model of WSe$_2$ shown to qualitatively reproduce the results of {\it ab-initio} time dependent density functional theory\cite{sharma_valley_2022,
sharma_thz_2023}. Full details of this model can be found in the Supplemental document.
In Fig.~\ref{fig3}(a,b) we present "maps" of the charge current and spin current for the two key parameters of few cycle circularly polarized light, the amplitude and full width half maximum.
Both the spin current, panel (a), and charge current, panel (b), are maximal at the single cycle and large amplitude regime. At pulse durations greater than 2.8~fs, i.e. in the few cycle regime, we see the appearance of nodal lines on which the current vanishes. These nodal lines, indicated by the arrows, result from exactly the multi-loop cancellation effects anticipated on the basis of the 2-band graphene model, and imply a rich spin current response: (i) a pure spin current, which occurs on the charge current nodal line; and (ii) pure charge currents, which occur on the on the spin current nodal line. This is revealed clearly in Fig.~\ref{fig3}(c) in which we plot the "spin purity", defined as

\begin{equation}
\eta_s = \frac{|\v J_Q|-|\v J_s|}{|\v J_Q|+|\v J_s|}
\label{P}
\end{equation}
that takes on values of -1 for a pure spin current, 0 for a 100\% spin polarized current, and +1 for a pure charge current. Regions of all of these cases, along with a continuous transition between them, may be seen in Fig.~\ref{fig3}(c).
Closer examination of the spin purity and spin and charge currents along the vertical line indicated in panel (c), reveal a pulse parameter tolerance for pure spin current of the order of $\pm 5$~\% to achieve spin purity $\eta_s > 0.8$, see Fig.~\ref{fig3}(d,e), with higher tolerances at larger FWHM although in this case the current amplitude is also weaker.

\begin{figure}[t!]
\includegraphics[width=0.49\textwidth]{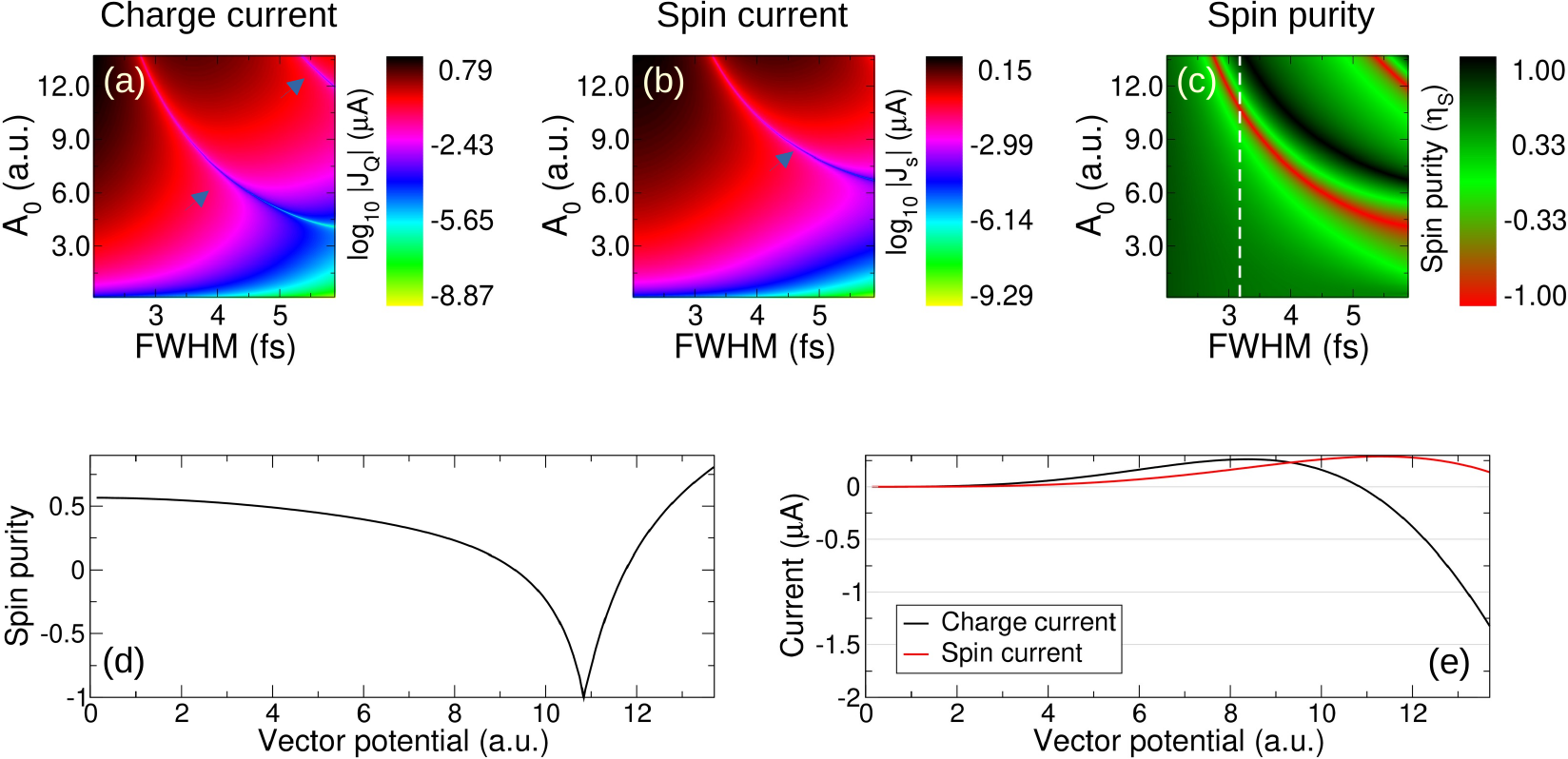}
\caption{{\it Spin and charge current response of WSe$_2$ in the symmetry breaking regime of circularly polarized light}. For gap tuned light of 2.25~eV variation of pulse full width half maximum (FWHM) and amplitude $A_0$ generates a charge current (a) and spin current (b) that exhibits nodes indicated by the arrows (note the scale in both cases is logarithmic). These nodal lines do not coincide, leading to a rich current response featuring pure spin current (the flow of spin current in the absence of charge flow), 100\% polarized spin currents, and pure charge currents, corresponding respectively to values of -1, 0, and +1 of the spin purity, Eq.~\ref{P}, shown in panel (c). Panels (d-e) exhibit the spin purity and spin and charge currents on the broken line displayed in panel (c).
}
\label{fig3}
\end{figure}

\begin{figure}[t!]
\includegraphics[width=0.49\textwidth]{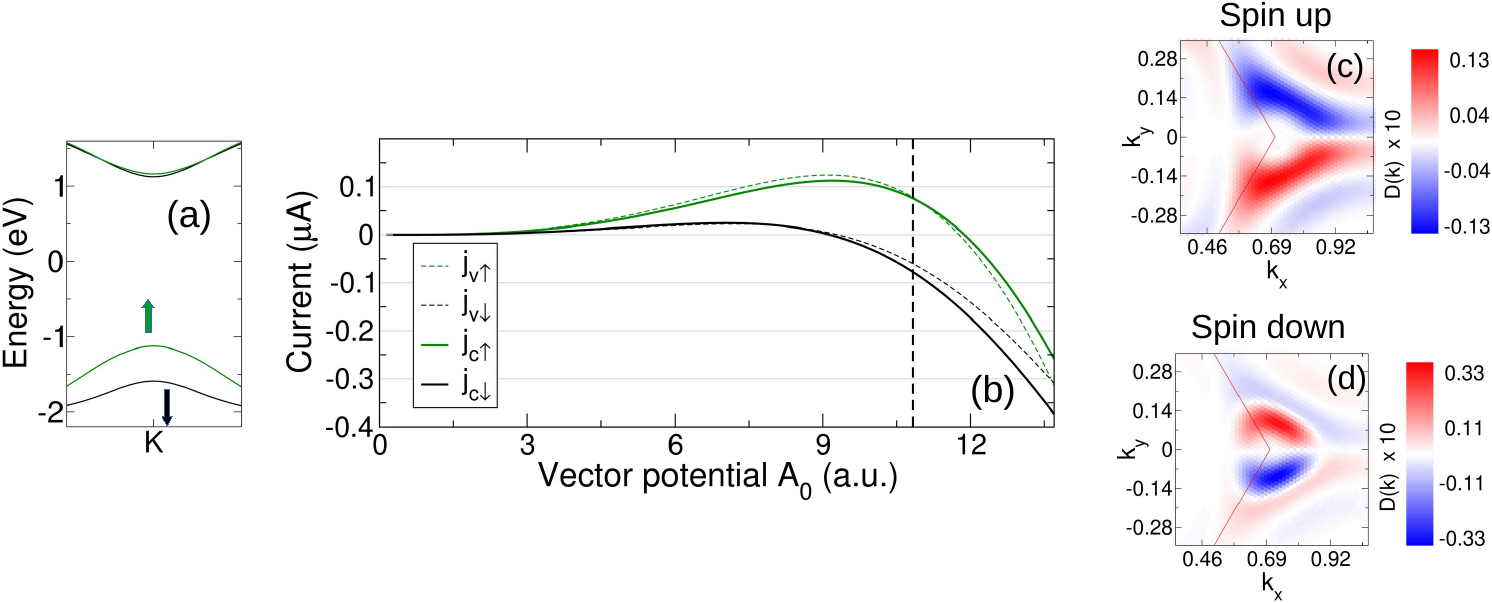}
\caption{{\it Mechanism of pure spin current generation}. A few cycle light pulse induces, for a tuned value of the vector potential, opposite currents in each spin channel yielding a spin flow but no net charge flow.
The valley band structure of WSe$_2$ possesses differing gaps in the spin up and down channels, panel (a), with the larger gap in the down channel at the K valley (with the opposite situation at the K$^\ast$ valley). (b) The light induced intraband current in each spin channel, which is seen to fall to zero at two different values of the vector potential amplitude $A_0$. The is, in consequence, a critical value of $A_0$, indicated by the vertical broken line, for which the two spin channels possess exactly opposite spin current. This corresponds to the absence of a net charge flow and thus to a pure spin current. Underpinning this is a reversal in orientation of the momentum space dipole structure that few cycle light excites at the K valley, as revealed by the $D(\v k)$ function, Eq.~\ref{DK}, plotted for the conduction spin up/down channels at the critical value of $A_0$, panels (c,d) respectively.
}
\label{fig4}
\end{figure}

To confirm the origin of this pure current  resides in the different response of the the spin up and down band manifolds to few cycle light at one valley, we analyse the current response of individual bands along the broken line shown in Fig.~\ref{fig3}(c-e). Due to the differing gaps in the two spin channels in WSe$_2$, see Fig.~\ref{fig4}(a), the turning point at which the current changes sign occurs at different vector potential amplitude in each band, Fig.~\ref{fig4}(b). As a result there occurs a critical value of the vector potential, indicated by the vertical line in Fig.~\ref{fig4}(b), at which both spin channels have finite but opposite current i.e.  a pure spin current.

Further insight into this current response can be obtained via the function

\begin{equation}
D(\v k) = |c_{\v k}|^2 - \frac{1}{3}
\sum_{i=1}^{3} |c_{M_i \v k}|^2.
\label{DK}
\end{equation}
that measures the deviation of the conduction band occupation at crystal momenta $\v k$, $|c_{\v k}|^2$, from the "star average" of the conduction band occupations averaged over the 3 $\v k$-vectors related by the valley $C_3$ symmetry operations\cite{sharma_direct_2024}. This thus acts as a "symmetry breaking density". At long pulse duration the dynamical trajectory in momentum space possesses approximate rotational symmetry and the excitation inherits the $C_3$ symmetry of the underlying valley manifold with $D(\v k)$ then zero for all $\v k$. However for short pulses, as shown in Fig.~\ref{fig1}(d), the dynamical trajectory possesses $C_2$ symmetry, lower that that of the underlying valley manifold, and the charge excitation inherits this lower symmetry, with $D(\v k)\ne 0$.

\begin{figure*}[t!]
\includegraphics[width=0.6\textwidth]{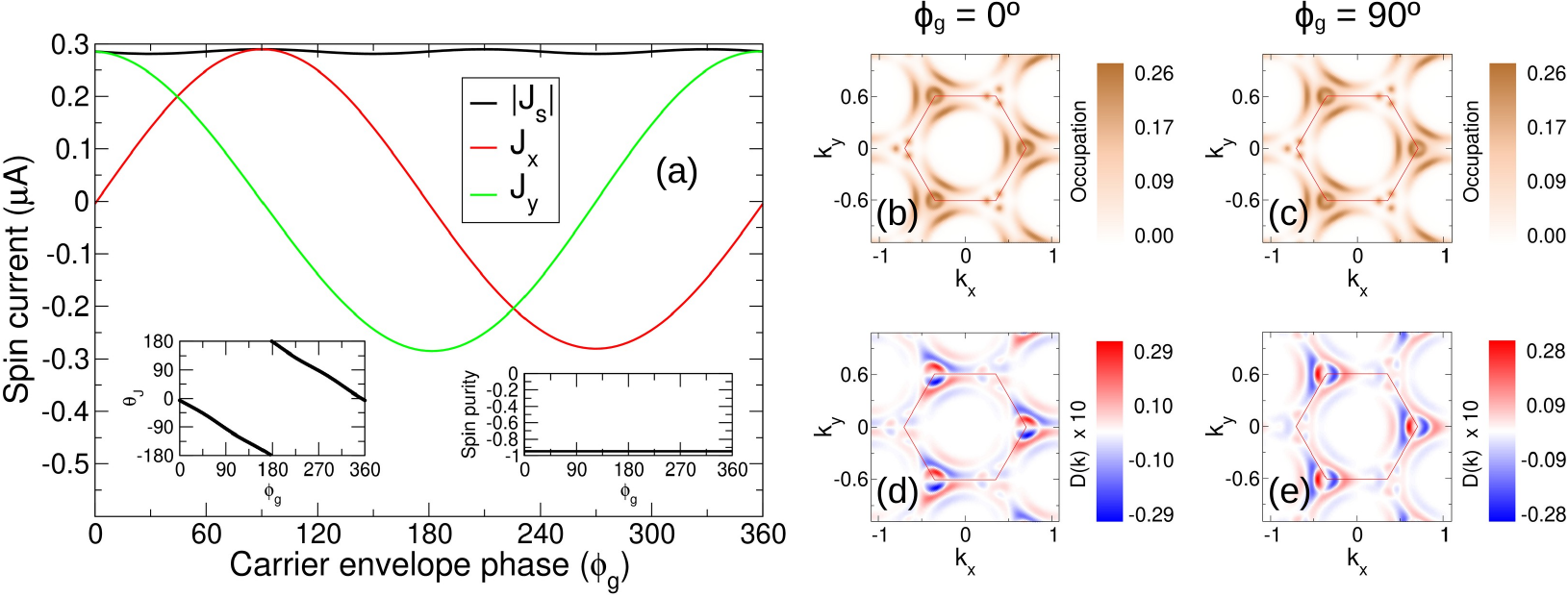}
\caption{{\it Light pulse control over pure spin current}. (a) The spin current as a function of the "global" carrier envelope phase $\phi_g$, see Eq.~\ref{phi}. The current direction is determined by this pulse parameter as $\theta_J = \pi - \phi_g$, see left inset panel. For all values of $\phi_g$ the spin purity remains close to $-1$ (i.e. a nearly perfect pure spin current), right hand inset panel. The momentum resolved excited state density, panels (b,c), and symmetry breaking density $D(\v k)$, Eq.~\ref{DK}, panels (d,e), are shown for two representative carrier envelope phases of $\phi_g = 0^\circ$ and $\phi_g = 90^\circ$. While the charge excitations appear similar for both angles, the $D(\v k)$ function clearly rotates with the changing pulse carrier envelope phase. 
}
\label{fig5}
\end{figure*}

This can be seen in Fig.~\ref{fig4}(c,d) in which plotted, at the critical vector potential amplitude $A_0$ corresponding to a pure spin current, $D(\v k)$ for the spin up and down conduction bands. One notes a dipole like structure at each valley centre with this dipole oriented oppositely in the two spin channels. This implies that in the spin up manifold there is more excited charge at $k_y < 0$ than at $k_y > 0$ (measured from the valley centre), and therefore a current flow $j_y < 0$, with the opposite situation in the spin down manifold. It is this low symmetry excitation, sensitive to the different gaps in each spin channel, that underpins the light induced pure spin current.

\emph{Light control over pure spin current:} The coupling of few cycle circularly polarized light to current implies an underlying vectorial degree of freedom of the pulse. A long time circularly polarized pulse evidently does not possess this: the continuously rotating polarization vector implies no fixed direction. However at the few cycle limit, the oscillation period and pulse envelope become of comparable duration and their relative temporal alignment important in the overall waveform. This is encoded via the angle $\phi_g$ in the vector pulse potential

\begin{equation}
\v A(t) = f(t) \left(\sigma \cos(\omega t + \phi_g),\sin(\omega t + \phi_g) \right).
\label{phi}
\end{equation}
with $f(t)$ the pulse envelope, $\sigma=\pm1$ the helicity, and $\omega$ the frequency. This can be thought of as a 'global' carrier envelope phase, in contrast to the relative carrier envelope phase of $\sigma\pi/2$ between the two pulse components required in circularly polarized light. As can be seen in Fig.~\ref{fig5}, the direction of the pure spin current is determined by this parameter, with the inset revealing that the current is aligned exactly along the angle $\phi_g$. The origin of this behaviour resides in the $C_2$ symmetry of the underlying momentum space trajectory induced by the light pulse, the axis of which is determined by $\phi_g$, and which in turn determines the direction of the $C_2$ symmetry broken excited state charge at the K valley. This can be seen in Fig.~\ref{fig5}(b-e) for two representative cases, $\phi_g=0^\circ$ and $\phi_g=90^\circ$, where in panels (b,c) is displayed the momentum resolved charge excitation and in panels (d,e) the symmetry lowering density $D(\v k)$, Eq.~\ref{DK}. While the charge excitations appear similar the $D(\v k)$ function, that records the $C_2$ symmetry lowering of the excited state charge, clearly exhibits rotation with the pulse global carrier envelope phase $\phi_g$.


{\it Discussion}: Many cycle pulses of circularly polarized light control spin-valley {charge} excitation: the well known spin-valley coupling that underpins lightwave spin- and valleytronics. Here we have shown that at the few cycle limit of such light pulses control emerges also over spin {current} physics. This ultrafast current response is extremely rich, with pure spin currents (the flow of spin without a net charge flow), 100\% spin polarized currents, and pure charge currents all accessible by tuning the pulse amplitude and duration in the few cycle limit.

The pure spin current, as it is created by a circularly polarized pulse, is also valley polarized (i.e. flows dominantly at one valley). This highlights a key difference in the present approach: it does not rely on a symmetry indued cancellation {\it between}  conjugate valleys, as explored extensively in photo-galvanic effect, but instead arises as a cancellation effect between different spin manifolds at a {\it single} valley. The current integrated over each spin manifold are exactly opposite, leading to zero net charge flow.
Few cycle light thus represents an "all optical" approach to generating pure current requiring no material property other than  spin split valley bands, evidently the minimal requirement to generate spin distinguished currents in the solid state.
Our work thus both highlights the wealth of control possibilities in the few cycle limit of light, as well as offering a potential route to an ultrafast charge and spin-current control in the transition metal dichalcogenides, including the generation of pure spin currents. 

\section{Ackcknowledgements}

Gill would like to thank DFG for funding through project-ID 328545488 TRR227 (project A04). Sharma would like to thank DFG for funding through project-ID 328545488 TRR227 (projects A04), and Shallcross would like to thank DFG for funding through project-ID 522036409 SH 498/7-1. The authors acknowledge the North-German Supercomputing Alliance (HLRN) for providing HPC resources that have contributed to the research results reported in this paper.



%



\clearpage            

\onecolumngrid 

\newpage




\section*{Supplementary Materials}

\begin{center}

\textbf{Generation of pure, spin polarized, and unpolarized charge currents at the few cycle limit of circularly polarized light}

Deepika~Gill,
Sangeeta~Sharma,
Sam~Shallcross$^{\ast}$\\
\small$^\ast$Corresponding author. Email: shallcross@mbi-berlin.de\\
\end{center}





\title[An \textsf{achemso} demo]
{Supplemental Information: Pure spin current generated by few cycle light: supplemental document}


	
	
	


	\section{Intraband and interband current}
	
	\begin{figure}[h!]
		\begin{center}
			\includegraphics[width=1\textwidth]{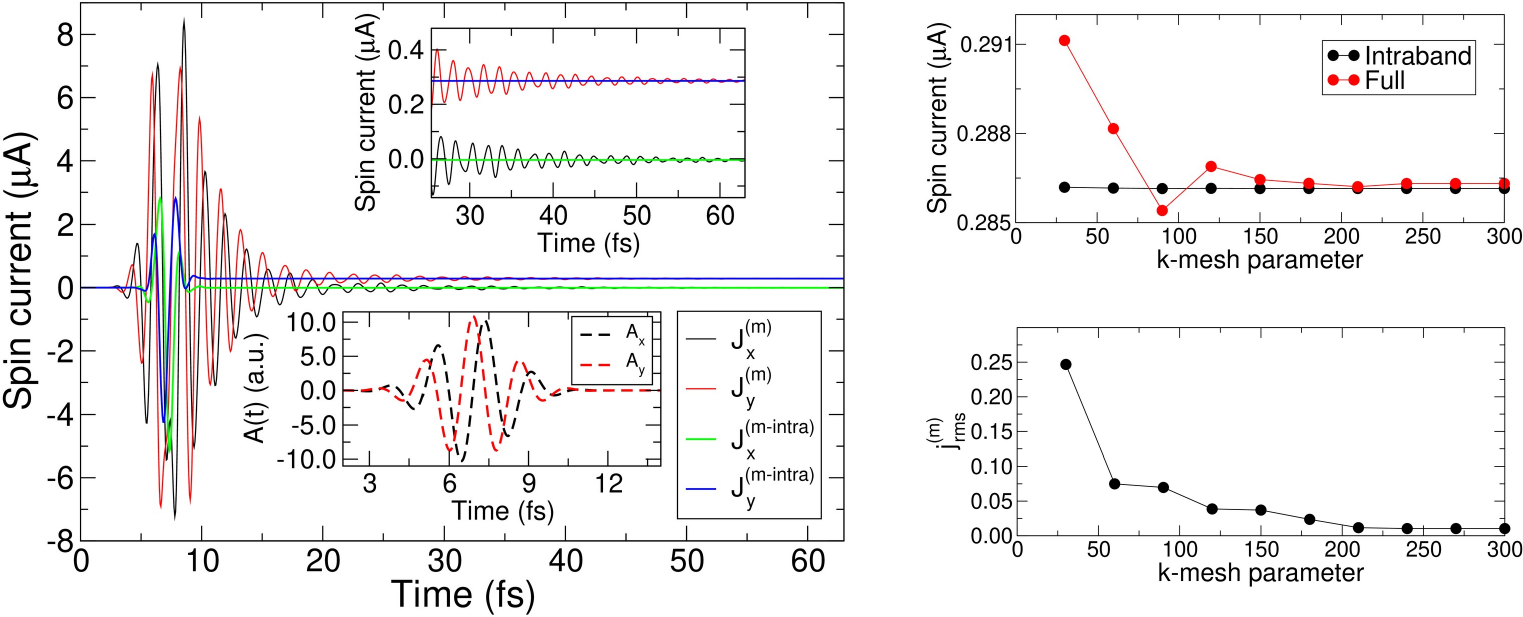}
			\caption{{\it Intra- and inter-band contribution to the light induced current}. (a) The spin current for the laser pulse shown in the lower inset panel (gap tuned frequency 2.25~eV, duration 3.17~fs, and amplitude 10.86~a.u.). Shown are both the intra-band component and the full current, with the upper right panel a zoom of the long time behaviour. Evidently, the full current converges in an oscillatory way to the intra-band component at sufficiently long time. The full current includes interference effects, that require careful convergence of the k-mesh, as shown in panels (b,c).
			}
			\label{S1}
		\end{center}
	\end{figure}
	
	The macroscopic current density at time $t$ can be written as
	
	\begin{equation}
		\v j(t) = \frac{1}{V_{UC}} \sum_{\v q} \mel{\Psi_{\v q}(t)}{\m\nabla_{\v q} H(\v q, t)}{\Psi_{\v q}(t)}
		\label{totj}
	\end{equation}
	where the sum $\v q$ is over k-vectors in the Brillouin zone of area $V_{UC}$ and $\ket{\Psi_{\v q}}$ the time dependent ket.
	
	The current can be decomposed in a number of physically useful ways, here we consider a separation into intra-band and inter-band contributions.
	Writing the time dependent ket $\ket{\Psi_{\v q}(t)}$ in the basis instantaneous eigenvectors at $\v k(t)$ as
	
	\begin{equation}
		\ket{\Psi_{\v q}(t)} = \sum_i c_{i \v q}(t) \ket{\Phi_{i\v k(t)}} 
	\end{equation}
	where the instantaneous eigenvalues and eigenvectors are defined by
	
	\begin{equation}
		H(\v k) \ket{\Phi_{i \v k}} = E_i(\v k) \ket{\Phi_{i \v k}}
	\end{equation}
	with the evolution of crystal momentum induced by the pulse, $\v k(t)$, given by the Bloch acceleration theorem
	
	\begin{equation}
		\v k(t) = \v q - \v A(t)/c,
		\label{BAC}
	\end{equation}
	we can decompose Eq.~\ref{totj} into two terms, intra- and inter-band contributions,
	
	\begin{equation}
		\v j(t) = \v j_{intra}(t) + \v j_{inter}(t)
	\end{equation}
	where
	
	\begin{equation}
		\v j_{intra}(t) = \frac{1}{V_{UC}} \sum_{i\v q} \left|c_{i\v q}\right|^2\m\nabla_{\v k} \left.E_{i}(\v k)\right|_{\v k = \v k(t)}
		\label{jintra}
	\end{equation}
	and
	
	\begin{equation}
		\v j_{inter}(t) = -\frac{1}{V_{UC}} \sum_{ij\v q} c_{i\v k(t)}^\ast c_{j\v k(t)} E_{j}(\v k(t)) \braket{\m\nabla_{\v p} \Phi_{i\v k(t)}}{\Phi_{j\v k(t)}} + c.c.
		\label{jinter}
	\end{equation}
	where we note that in both these expressions the sum over $\v q$ enters through $\v k(t)$ as given by the Bloch acceleration theorem, Eq.~\ref{BAC}.
	
	A key difference between these two contributions is that while intra-band contribution is determined solely by occupation numbers, the inter-band term, Eq.~\ref{jinter}, evidently involves interference between valence and conduction states. This generates an oscillatory contribution that decreases with increasing time (as at large times the dynamical phases change increasingly quickly with $\v k$ leading to cancellation when integrated over the Brillouin zone.
	
	This cancellation requires significantly finer k-grids to capture than the intra-band term, a fact shown in Fig.~\ref{S1}. The main panel (a) displays the $x$- and $y$-components of the total spin current, $J^{(m)}_{x,y}$, alongside the intra-band component of this current, $J^{(m-intra)}_{x,y}$. The full current is seen to converge with increasing time to the intra-band component, with the upper inset panel a zoom in of the long time behaviour. The lower inset panel displays the laser pulse that has a vector potential amplitude $A_0 = 10.86$~a.u., a full width half maxima of 3.17~fs, and a gap tuned frequency of 2.25~eV. Panels (b,c) display the convergence with k-mesh parameter $n_k$ describing an $n_k \times n_k$ mesh; evidently while the intra-band current converges very rapidly, the full current exhibits a slow convergence due to the inter-band component.
	
	\section{Tight-binding methodology}
	
	\begin{table}[t!]
		\centering
		\begin{tabular}{|c|c|c|c|c|}  \hline
			$\Delta$ & $\Delta_{v}^{SO}$ & $\Delta_{c}^{SO}$ & $t_{\bot}$ & $\lambda_{SO}$  \\ \hline
			2.5 &  0.466 eV   &  -0.037 eV & -1.4 eV  &  0.05 \\ \hline
		\end{tabular}
		\caption{Parameters for calculations of WSe$_2$ in a minimal four-band model.}
		\label{SOtab}
	\end{table}
	
	{\it Laser pulse}: The laser pulse components in all the calculations is characterised by a waveform composed of a Gaussian envelope that modulates a sinusoidal oscillation centred at the pulse time $t_0^{(i)}$, with the full vector potential is then obtained by summing over such components:
	
	\begin{equation}
		\v A(t) = \sum_i \v A_0^{(i)} \exp\left(-\frac{(t-t_0^{(i)})^2}{2\sigma_i^2}\right) \cos(\omega_i t + \phi_{cep}^{(i)})
		\label{pulse}
	\end{equation}
	where $\v A_0^{(i)}$ is the polarization vector with magnitude $A_0^{(i)}=|\v A_0^{(i)}|$ and polarization direction 
	$\hat{\v A}^{(i)}_0=\v A_0^{(i)}/A_0^{(i)}$. 
	The remaining pulse characteristics are $\omega_i$, the central frequency of the pulse component, $\phi_{cep}^{(i)}$, the carrier envelop phase, and $\sigma_i$ which is related to the full width half maximum (FWHM) of the pulse through $2\sqrt{2\ln{2}}\sigma_i$.
	
	{\it Dynamics}: The density matrix evolves according to the von Neumann equation, which we solve using a basis composed of the underlying Hamiltonian eigenstates. At time $t$ these eigenstates are evaluated at crystal momentum $\v k(t)$, given by the Bloch acceleration theorem
	
	\begin{equation}
		\v k(t) = \v k(0) - \v A(t)/c.
	\end{equation}
	To address quantum decoherence of the density matrix $\rho$, we adopt a straightforward phenomenological approach, which involves an exponential decay of the off-diagonal elements of $\rho$. As the density matrix is represented in the eigenbasis at $\v k(t)$, these off-diagonal elements encapsulate quantum interference effects, and their decay the suppression of quantum coherence. The resulting von Neumann equation can be written as
	
	\begin{equation}
		\partial_t \rho = -i\left[H,\rho\right] + \frac{1}{T_{2}} (\rho-\text{Diag}[\rho])
		\label{LVN}
	\end{equation}
	where $T_{2}$ is a phenomenological decoherence time that determines the timescale for the attenuation of interference effects, and $\text{Diag}[\rho]$ the matrix that contains only the diagonal components of the density matrix $\rho$. We use the standard fourth order Runge-Kutta method for numerical time evolution.
	
	{\it Gapped graphene}: The tight-binding Hamiltonian for gapped graphene can be written in sub-lattice space as
	
	\begin{equation}
		H_{\sigma} = \begin{pmatrix}
			\Delta & f_{\v k} \\ f^\ast_{\v k} & -\Delta \end{pmatrix}
	\end{equation}
	where the Bloch sum $f_k = -\sum_j t e^{ik.\nu_j}$ where $j$ runs over nearest neighbopur vectors $\m \nu_j$, with $t$ the nearest neighbour hopping and $\v k$ the crystal momentum. The field $\Delta$ represents an on-site field that alternates sign between the two sublattices. 
	
	{\it Tight-binding model employed for WSe$_2$}: In order to fully understand the dynamic development of a material, it is necessary to have a complete Brillouin zone band structure. This means that we cannot only focus on the low energy band structure in the vicinity of high symmetry K points, which is typically done with the $\v k.\v p$ technique~\cite{kormanyos2015k}. Rather, we use a simple model that describes the two low-energy spin states with a gapped graphene-based model~\cite{sharma2023thz}. Within the low-energy regime, spin-orbit (SO) coupling is diagonal in both sub-lattice and spin space and can be written as $\lambda_{SO} \sigma_z \otimes \tau_z$ ($\sigma_z$ denotes Pauli matrices in spin space and $\tau_z$ Pauli matrices in sub-lattice space). This approach is justified as it allows us to obtain the low-energy Hamiltonian by including both the band and SO splitting terms.
	
	\begin{equation}
		H = \begin{pmatrix}
			H_\uparrow & 0 \\ 0 & H_\downarrow 
		\end{pmatrix}
		\label{HW0}
	\end{equation}
	where
	
	\begin{equation}
		H_{\sigma} = \begin{pmatrix}
			\Delta & f_{\v k} \\ f^\ast_{\v k} & -\Delta \end{pmatrix}
		+
		\sigma
		\begin{pmatrix}
			\Delta_v^{SO} & 0 \\ 0 & \Delta_c^{SO}
		\end{pmatrix}
		f_{SO}(\v k)
		\label{HW}
	\end{equation}
	where
	
	\begin{equation}
		f_k = \sum_jt_{\bot}e^{ik.\nu_j}
		\label{FK}
	\end{equation}
	and where $\nu_j$ represent the nearest-neighbour vectors, and the SO scale function is given by 
	
	\begin{equation}
		f_{SO}(k) = \sum_{k_{MT}}\nu(k_{MT})\exp[-\frac{1}{2}(|k| - |k_{MT}|)^2/\lambda_{SO}^2] 
	\end{equation}
	In this expression $\nu(k_{MT})$ takes on values of +1 at K valley and -1 and the K$^*$ valley and the sum $k_{MT}$ is over the union of the translation groups of the two inequivalent K and K$^*$ points. The value of $\lambda_{\textrm{SO}}$ can be found in Table 1 and is chosen so that $f_{SO}$ (k) falls to zero outside the vicinity of the low-energy K valleys. 
	
	The SO splitting in WSe$2$ are $\Delta_v^{SO}=0.466$~eV for the valence band and $\Delta_c^{SO}=-0.037$~eV for the conduction band\cite{kormanyos_kptheory_2015}, and employing the values given in Table~\ref{SOtab} yields these SO splitting with a band gap of 2.25~eV.
	
	{\it Tight-binding employed for bilayer graphene}: We employ a standard nearest neighbour minimal $\pi$-band model for bilayer graphene\cite{mccann_electronic_2013} with in-plane and interlayer hopping parameters $t=-3.2$~eV and $t_\perp = 0.4$~eV respectively. The bilayer graphene Hamiltonian is then given by
	
	\begin{equation}
		H_{BLG} = \begin{pmatrix}
			0 & t_{\v k} & t_\perp & 0 \\
			t_{\v k}^\ast & 0 & 0 & 0 \\
			t_\perp & 0 & 0 & t_{\v k}^\ast \\
			0 & 0 & t_{\v k} & 0 
		\end{pmatrix}
	\end{equation}
	
	
	
	\newpage
	\section{Switching the orientation of the K-pole}
	
	\begin{figure}[h!]
		\begin{center}
			\includegraphics[width=1\textwidth]{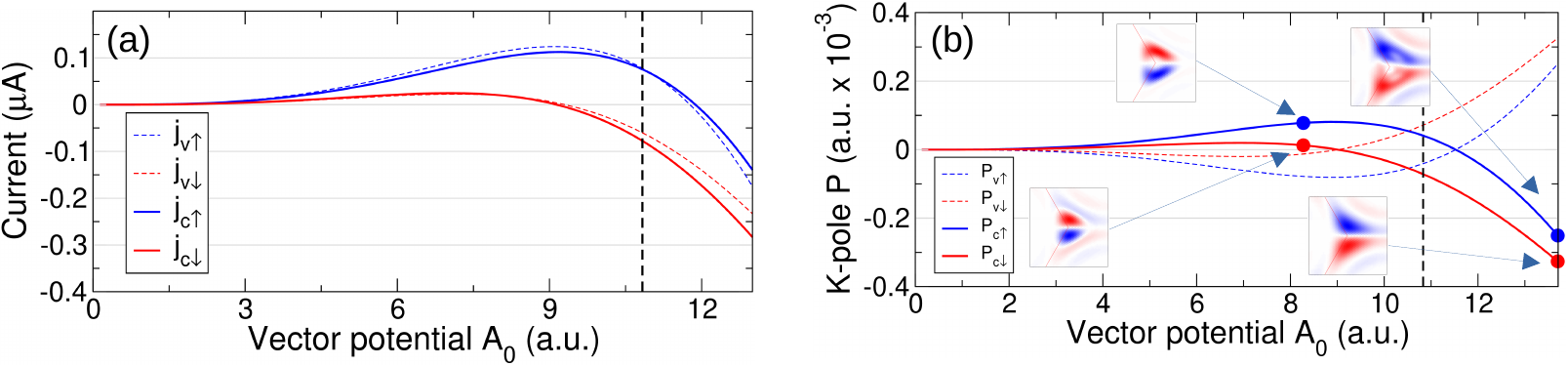}
			\caption{{\it Switching of the direction of the momentum space dipole at the K point ("K-pole") with increasing vector potential}. (a) The current changes direction with increasing vector potential amplitude $A_0$ (the current is aligned along the $y$-direction for all $A_0$ hence can be represented by a signed scalar). This change in sign is driven by a change in the symmetry breaking of the excited state distribution in the K valley conduction band. The $D(\v k)$ function (see main text for details) shows a clear reversal with increasing $A_0$ with the dipole moment calculated from this distribution, panel (b), showing a change of sign exactly where the current, panel (a), changes sign.
			}
			\label{S2}
		\end{center}
	\end{figure}
	
	%

		%
	%
%
	%
		
	
\end{document}





\maketitle

\section{Intraband and interband current}

\begin{figure}[h!]
\begin{center}
\includegraphics[width=1\textwidth]{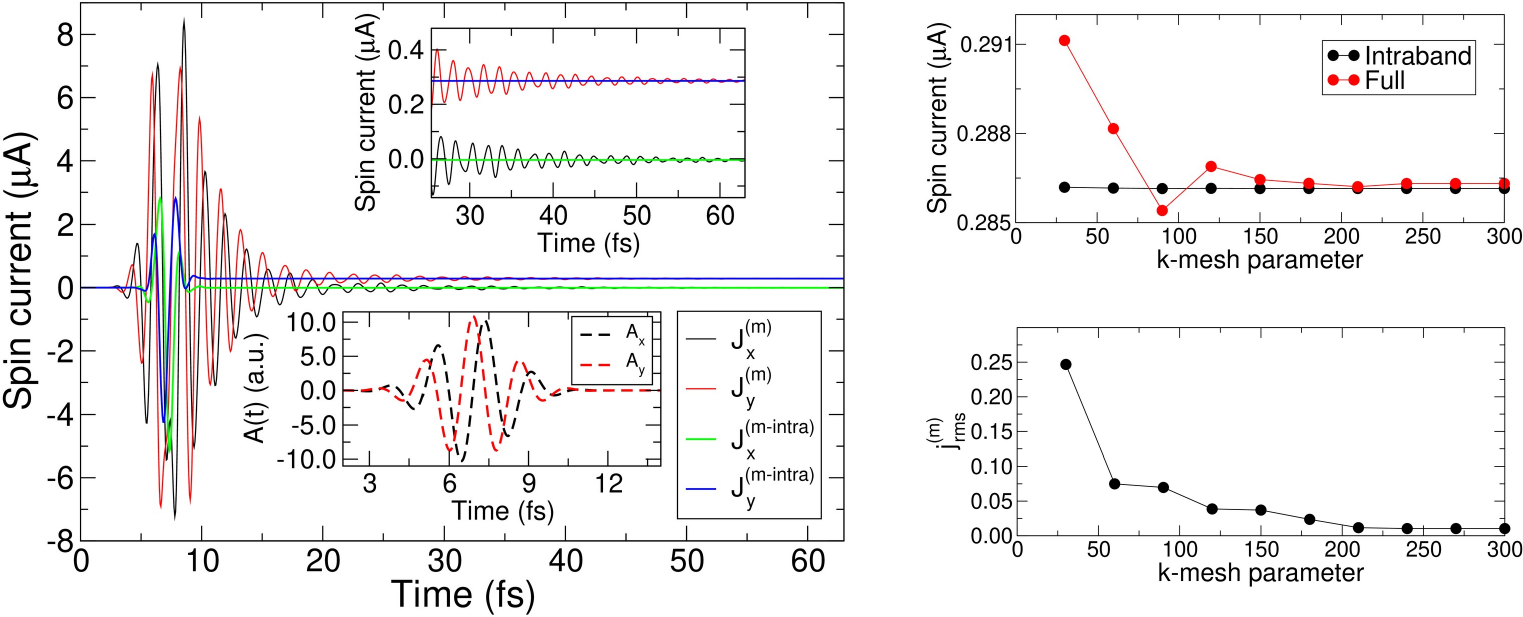}
\caption{{\it Intra- and inter-band contribution to the light induced current}. (a) The spin current for the laser pulse shown in the lower inset panel (gap tuned frequency 2.25~eV, duration 3.17~fs, and amplitude 10.86~a.u.). Shown are both the intra-band component and the full current, with the upper right panel a zoom of the long time behaviour. Evidently, the full current converges in an oscillatory way to the intra-band component at sufficiently long time. The full current includes interference effects, that require careful convergence of the k-mesh, as shown in panels (b,c).
}
\label{S1}
\end{center}
\end{figure}

The macroscopic current density at time $t$ can be written as

\begin{equation}
\v j(t) = \frac{1}{V_{UC}} \sum_{\v q} \mel{\Psi_{\v q}(t)}{\m\nabla_{\v q} H(\v q, t)}{\Psi_{\v q}(t)}
\label{totj}
\end{equation}
where the sum $\v q$ is over k-vectors in the Brillouin zone of area $V_{UC}$ and $\ket{\Psi_{\v q}}$ the time dependent ket.

The current can be decomposed in a number of physically useful ways, here we consider a separation into intra-band and inter-band contributions.
Writing the time dependent ket $\ket{\Psi_{\v q}(t)}$ in the basis instantaneous eigenvectors at $\v k(t)$ as

\begin{equation}
\ket{\Psi_{\v q}(t)} = \sum_i c_{i \v q}(t) \ket{\Phi_{i\v k(t)}} 
\end{equation}
where the instantaneous eigenvalues and eigenvectors are defined by

\begin{equation}
H(\v k) \ket{\Phi_{i \v k}} = E_i(\v k) \ket{\Phi_{i \v k}}
\end{equation}
with the evolution of crystal momentum induced by the pulse, $\v k(t)$, given by the Bloch acceleration theorem

\begin{equation}
\v k(t) = \v q - \v A(t)/c,
\label{BAC}
\end{equation}
we can decompose Eq.~\ref{totj} into two terms, intra- and inter-band contributions,

\begin{equation}
\v j(t) = \v j_{intra}(t) + \v j_{inter}(t)
\end{equation}
where

\begin{equation}
\v j_{intra}(t) = \frac{1}{V_{UC}} \sum_{i\v q} \left|c_{i\v q}\right|^2\m\nabla_{\v k} \left.E_{i}(\v k)\right|_{\v k = \v k(t)}
\label{jintra}
\end{equation}
and

\begin{equation}
\v j_{inter}(t) = -\frac{1}{V_{UC}} \sum_{ij\v q} c_{i\v k(t)}^\ast c_{j\v k(t)} E_{j}(\v k(t)) \braket{\m\nabla_{\v p} \Phi_{i\v k(t)}}{\Phi_{j\v k(t)}} + c.c.
\label{jinter}
\end{equation}
where we note that in both these expressions the sum over $\v q$ enters through $\v k(t)$ as given by the Bloch acceleration theorem, Eq.~\ref{BAC}.

A key difference between these two contributions is that while intra-band contribution is determined solely by occupation numbers, the inter-band term, Eq.~\ref{jinter}, evidently involves interference between valence and conduction states. This generates an oscillatory contribution that decreases with increasing time (as at large times the dynamical phases change increasingly quickly with $\v k$ leading to cancellation when integrated over the Brillouin zone.

This cancellation requires significantly finer k-grids to capture than the intra-band term, a fact shown in Fig.~\ref{S1}. The main panel (a) displays the $x$- and $y$-components of the total spin current, $J^{(m)}_{x,y}$, alongside the intra-band component of this current, $J^{(m-intra)}_{x,y}$. The full current is seen to converge with increasing time to the intra-band component, with the upper inset panel a zoom in of the long time behaviour. The lower inset panel displays the laser pulse that has a vector potential amplitude $A_0 = 10.86$~a.u., a full width half maxima of 3.17~fs, and a gap tuned frequency of 2.25~eV. Panels (b,c) display the convergence with k-mesh parameter $n_k$ describing an $n_k \times n_k$ mesh; evidently while the intra-band current converges very rapidly, the full current exhibits a slow convergence due to the inter-band component.

\section{Tight-binding methodology}

\begin{table}[t!]
    \centering
    \begin{tabular}{|c|c|c|c|c|}  \hline
        $\Delta$ & $\Delta_{v}^{SO}$ & $\Delta_{c}^{SO}$ & $t_{\bot}$ & $\lambda_{SO}$  \\ \hline
         2.5 &  0.466 eV   &  -0.037 eV & -1.4 eV  &  0.05 \\ \hline
    \end{tabular}
    \caption{Parameters for calculations of WSe$_2$ in a minimal four-band model.}
    \label{SOtab}
\end{table}

{\it Laser pulse}: The laser pulse components in all the calculations is characterised by a waveform composed of a Gaussian envelope that modulates a sinusoidal oscillation centred at the pulse time $t_0^{(i)}$, with the full vector potential is then obtained by summing over such components:

\begin{equation}
\v A(t) = \sum_i \v A_0^{(i)} \exp\left(-\frac{(t-t_0^{(i)})^2}{2\sigma_i^2}\right) \cos(\omega_i t + \phi_{cep}^{(i)})
\label{pulse}
\end{equation}
where $\v A_0^{(i)}$ is the polarization vector with magnitude $A_0^{(i)}=|\v A_0^{(i)}|$ and polarization direction 
$\hat{\v A}^{(i)}_0=\v A_0^{(i)}/A_0^{(i)}$. 
The remaining pulse characteristics are $\omega_i$, the central frequency of the pulse component, $\phi_{cep}^{(i)}$, the carrier envelop phase, and $\sigma_i$ which is related to the full width half maximum (FWHM) of the pulse through $2\sqrt{2\ln{2}}\sigma_i$.

{\it Dynamics}: The density matrix evolves according to the von Neumann equation, which we solve using a basis composed of the underlying Hamiltonian eigenstates. At time $t$ these eigenstates are evaluated at crystal momentum $\v k(t)$, given by the Bloch acceleration theorem

\begin{equation}
\v k(t) = \v k(0) - \v A(t)/c.
\end{equation}
To address quantum decoherence of the density matrix $\rho$, we adopt a straightforward phenomenological approach, which involves an exponential decay of the off-diagonal elements of $\rho$. As the density matrix is represented in the eigenbasis at $\v k(t)$, these off-diagonal elements encapsulate quantum interference effects, and their decay the suppression of quantum coherence. The resulting von Neumann equation can be written as

\begin{equation}
\partial_t \rho = -i\left[H,\rho\right] + \frac{1}{T_{2}} (\rho-\text{Diag}[\rho])
\label{LVN}
\end{equation}
where $T_{2}$ is a phenomenological decoherence time that determines the timescale for the attenuation of interference effects, and $\text{Diag}[\rho]$ the matrix that contains only the diagonal components of the density matrix $\rho$. We use the standard fourth order Runge-Kutta method for numerical time evolution.

{\it Gapped graphene}: The tight-binding Hamiltonian for gapped graphene can be written in sub-lattice space as

\begin{equation}
H_{\sigma} = \begin{pmatrix}
\Delta & f_{\v k} \\ f^\ast_{\v k} & -\Delta \end{pmatrix}
\end{equation}
where the Bloch sum $f_k = -\sum_j t e^{ik.\nu_j}$ where $j$ runs over nearest neighbopur vectors $\m \nu_j$, with $t$ the nearest neighbour hopping and $\v k$ the crystal momentum. The field $\Delta$ represents an on-site field that alternates sign between the two sublattices. 

{\it Tight-binding model employed for WSe$_2$}: In order to fully understand the dynamic development of a material, it is necessary to have a complete Brillouin zone band structure. This means that we cannot only focus on the low energy band structure in the vicinity of high symmetry K points, which is typically done with the $\v k.\v p$ technique~\cite{kormanyos2015k}. Rather, we use a simple model that describes the two low-energy spin states with a gapped graphene-based model~\cite{sharma2023thz}. Within the low-energy regime, spin-orbit (SO) coupling is diagonal in both sub-lattice and spin space and can be written as $\lambda_{SO} \sigma_z \otimes \tau_z$ ($\sigma_z$ denotes Pauli matrices in spin space and $\tau_z$ Pauli matrices in sub-lattice space). This approach is justified as it allows us to obtain the low-energy Hamiltonian by including both the band and SO splitting terms.

\begin{equation}
H = \begin{pmatrix}
H_\uparrow & 0 \\ 0 & H_\downarrow 
\end{pmatrix}
\label{HW0}
\end{equation}
where

\begin{equation}
H_{\sigma} = \begin{pmatrix}
\Delta & f_{\v k} \\ f^\ast_{\v k} & -\Delta \end{pmatrix}
+
\sigma
\begin{pmatrix}
\Delta_v^{SO} & 0 \\ 0 & \Delta_c^{SO}
\end{pmatrix}
f_{SO}(\v k)
\label{HW}
\end{equation}
where

\begin{equation}
    f_k = \sum_jt_{\bot}e^{ik.\nu_j}
\label{FK}
\end{equation}
and where $\nu_j$ represent the nearest-neighbour vectors, and the SO scale function is given by 

\begin{equation}
    f_{SO}(k) = \sum_{k_{MT}}\nu(k_{MT})\exp[-\frac{1}{2}(|k| - |k_{MT}|)^2/\lambda_{SO}^2] 
\end{equation}
In this expression $\nu(k_{MT})$ takes on values of +1 at K valley and -1 and the K$^*$ valley and the sum $k_{MT}$ is over the union of the translation groups of the two inequivalent K and K$^*$ points. The value of $\lambda_{\textrm{SO}}$ can be found in Table 1 and is chosen so that $f_{SO}$ (k) falls to zero outside the vicinity of the low-energy K valleys. 

The SO splitting in WSe$2$ are $\Delta_v^{SO}=0.466$~eV for the valence band and $\Delta_c^{SO}=-0.037$~eV for the conduction band\cite{kormanyos_kptheory_2015}, and employing the values given in Table~\ref{SOtab} yields these SO splitting with a band gap of 2.25~eV.

{\it Tight-binding employed for bilayer graphene}: We employ a standard nearest neighbour minimal $\pi$-band model for bilayer graphene\cite{mccann_electronic_2013} with in-plane and interlayer hopping parameters $t=-3.2$~eV and $t_\perp = 0.4$~eV respectively. The bilayer graphene Hamiltonian is then given by

\begin{equation}
H_{BLG} = \begin{pmatrix}
0 & t_{\v k} & t_\perp & 0 \\
t_{\v k}^\ast & 0 & 0 & 0 \\
t_\perp & 0 & 0 & t_{\v k}^\ast \\
0 & 0 & t_{\v k} & 0 
\end{pmatrix}
\end{equation}



\newpage
\section{Switching the orientation of the K-pole}

\begin{figure}[h!]
\begin{center}
\includegraphics[width=1\textwidth]{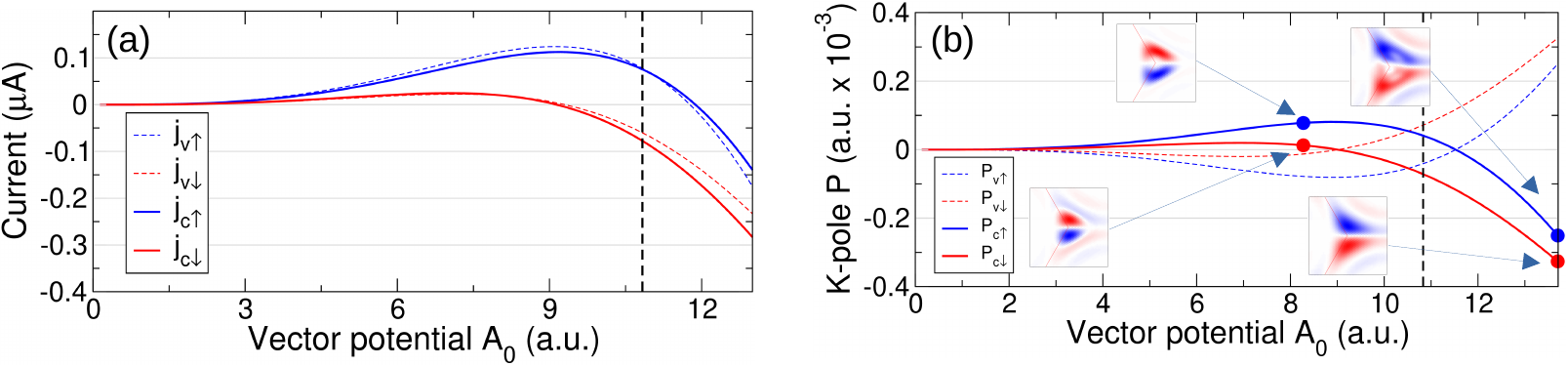}
\caption{{\it Switching of the direction of the momentum space dipole at the K point ("K-pole") with increasing vector potential}. (a) The current changes direction with increasing vector potential amplitude $A_0$ (the current is aligned along the $y$-direction for all $A_0$ hence can be represented by a signed scalar). This change in sign is driven by a change in the symmetry breaking of the excited state distribution in the K valley conduction band. The $D(\v k)$ function (see main text for details) shows a clear reversal with increasing $A_0$ with the dipole moment calculated from this distribution, panel (b), showing a change of sign exactly where the current, panel (a), changes sign.
}
\label{S2}
\end{center}
\end{figure}
